\documentclass[12pt]{article}
\usepackage{latexsym,amsmath,amssymb,amsbsy,graphicx}

\title{{\bf Tests of cryogenic FP-cavity with mirrors on different substrates.}}
\author{M.V.~Kuvshinskii$^1$, S.I.~Oreshkin$^1$, S.M.~Popov$^1$,\\ 
V.N.~Rudenko$^{1,a}$, I.S.~Yudin$^{1,2}$ and V.V.~Azarova$^3$, S.V.~Blagov$^4$\\
\\
\small $^{1}$ Sternberg Astronomical Institute, Lomonosov Moscow State University,\\
\small Universitetskii pr.13, Moscow 119992, Russia \\
\small $^{2}$ Moscow Physics Technical Institute \,(MPTI),\\
\small Dolgoprudnyi town,Institutskii pr.9, Moscow region 141701, Russia \\
\small $^{3}$ Polyus Stelmakh Research Institute,\\
\small Vvedenskogo St. 3, Moscow , 117342, Russia\\
\small $^{4}$ Moscow Research Television Institute, (MRTI)\\
\small Golianov St. 7a, Moscow , 115094, Russia\\
\\
\small $^a$ valentin.rudenko@gmail.com
}

\begin{document}
\maketitle

\begin{abstract}
Experiments were performed with Fabry-Perot optical resonators in vacuum at low temperatures. Mirrors were applied on substrates of various optical materials. The infrared laser with a wavelength of 1.064 microns was used. The pump power at the maximum could reach 450 mW. The evolution of the optical properties of the FP cavity was traced in the temperature range $(300-10)^o$K. The main parameters measured were the integral characteristics of the FP resonances - sharpness (finesse) and contrast of interference. Three types of substrates were tested: a sitall - the optical glass with ultra low thermal expansion (ULE), sapphire and calcium fluoride. During cooling, the degradation of the integral characteristics of the FP cavity was observed for the sitall mirrors due to the loss of the properties of (ULE), for sapphire mirrors due to the birefringence effect. The satisfactory constancy of the integral characteristics of the FP resonators on calcium fluoride was demonstrated in the entire temperature range studied 
\end{abstract}


\section{Introduction}
At present, optical Fabry-Pero cavities are used for different applications and for precise measurements in fundamental physics. In particular, they are used to address the problem of optical frequency standards \cite{bm-lud} and for gravitational wave detection using large-scale laser interferometers \cite{bm-rana,bm-will}.

Atomic transitions used as reference frequencies in optical clocks usually have a quality factor in the region of $10^{15}$, which corresponds to a line width of about 1 Hz or less. To explore such narrow transitions, an optical local oscillator is required that has a similar or narrower line width. Available solid-state and semiconductor diode lasers have a line width of tens of kHz to tens of MHz. Reducing the laser line width to the required level is most often achieved using the Pound-Drever-Hall stabilization method \cite{bm-drever}, which binds the optical generator to the ultra-stable optical Fabry-Perot cavity with a narrow line (high finesse). Such reference resonator (spacer) used for this purpose consists of two concave mirrors that are attached to each end of a solid-state hollow cylindrical specimen by optical contact. Both the sample and the mirror substrates are usually made of ultra-low thermal expansion glass (ULE), such that the coefficient of thermal expansion is of the order $(3\pm 0.3) \cdot 10^{-7}\, ^{o}$C in the temperature range of $(5-40)^o$C, crossing zero near room temperature. The typical size of the reference resonator (spacer) is about 10 cm in length with a high value of the number of reflected rays, finesse, of the order of $10^{5}$, which corresponds to the width of the optical resonator line in the range 5--10~kHz.

With good temperature control at the point of zero thermal expansion and suppression of the influence of mechanical vibrations, the frequency stability of the reference optical resonator (spacer) is limited only by dimensional changes due to thermal vibrations of the substrates of mirrors and their coatings \cite{bm-num}.
 
Currently, ULE glasses are used for optical reference resonators, since they have zero expansion near room temperature. A significant reduction in thermal noise can be achieved by switching to silica, which has much lower mechanical losses. Then the thermal noise will be determined mainly by coating (sprayed) and will be at a lower level. However, silica has a relatively large thermal expansion, and therefore very good temperature control is required.

An alternative would be to use a body of a reference resonator (spacer) made of ULE glass with low thermal noise and silica for substrates of mirrors \cite{bm-not} . Mirror coatings are alternating layers of silica (SiO$_2$) and tantalum (Ta$_2$O$_5$), with tantalum layers making the largest contribution to thermal noise \cite{bm-pen}. The cooling of the cavity itself will reduce the thermal noise, since it decreases as the square root of the temperature. However, in the case of a ULE, this would mean working away from the zero point of thermal expansion, which can lead to unacceptably high drift velocities. Silicon with zero expansion at $\sim$~120~$^{o}$K \cite{bm-oka} was proposed as an alternative material. Nevertheless, the problem of noise in the coating films remains. It becomes even more serious at low temperature, since the coefficient of mechanical loss of tantalum has a resonance at $^{o}K$ \cite{bm-franc}. In addition, silicon is transparent only for wavelengths above 1.2~microns. Currently, by scientific groups involved in the detection of gravitational waves, a significant work is being done to develop coatings with lower losses, and, for example, a halving was achieved by doping tantalum layers with titanium (TiO$_2$) \cite{bm-har}.

The prospects for increasing the sensitivity of laser interferometric gravitational-wave antennas \cite{bm-punt,bm-ligo} of the third generation are closely related to the idea of suppressing thermal noise of mirrors (test bodies) through their deep cooling to liquid nitrogen or helium temperature \cite{bm-som}. The pioneers of this direction were the Japanese groups who put forward the project of the first gravitational-wave interferometer with cryogenic mirrors \cite{bm-kur} at the end of the 20th century. This project was then tested in the pilot version \cite{bm-yam}, and is currently implemented as an underground cryogenic interferometer KAGRA \cite{bm-kag1,bm-kag2}.

To combat thermal noise, quantum photon noise and light pressure noise in a KAGRA setup, large heavy sapphire mirrors on sapphire suspensions \cite{bm-uchi} with low mechanical losses \cite{bm-cole} in the coating films were used. At the same time, mirrors weighing tens of kilograms working under the influence of powerful optical pumping are suspended as pendulums for effective isolation from the background of seismic noise. This is required to preserve the high mechanical quality facture of mirrors and suspensions. Under such conditions, the problem of cooling mirrors to low temperatures becomes a serious problem due to poor thermal contact. In particular, when cooling suspended quartz mirrors through the exchange gas in the presence of a luminous flux of about 20 mW, the temperature of about $(12-14)\, ^o$K was reached during a month \cite{bm-agat}. In contrast to the reference FP cavities required in the technique of optical frequency standards, the finesse of gravitational-wave interferometers with a kilometer base are much low (no more than 500), but their mirrors must withstand the high-power optical radiation (tens of kilowatts or more). Such requirements are also laid down in the European project Einstein Telescope (ET) \cite{bm-punt} and in the development projects of third-generation interferometers LIGO \cite{bm-ligo}. LIGO plans to switch to silicon mirrors with cooling to 120~$^o$K \cite{bm-amico}. The ET project is still considering this possibility along with other options \cite{bm-hild}. 

The search continues for various materials for test mass mirrors and their spectrally selective reflective coatings. Their evolution under conditions of long-term operation in gravitational-wave bursts monitoring is investigated.

In a recent paper by the KAGRA \cite{bm-has} group, the effect of changing the reflectivity of sapphire mirrors due to the adsorption of residual gas molecules (H2O molecules) on the cooled surface of mirrors was studied. An auxiliary measuring FP resonator with a finesse of about 50~000 was used. This increased finesse is required to register small changes in the reflectivity of the mirrors as a result of adsorption. The evaluation of the finesse was carried out by measuring the optical relaxation time of the Fabry-Perot resonance when the optical power was turned off.

In the observation interval of 35 days with a vacuum of $6\cdot 10^{-6}$~Pa, a drop in the finesse value of up to 5\% recorded (the growth rate of the adsorbed layer was estimated to be $27\pm 1.9$~nm/day). In future gravitational-wave interferometers of the third generation, this effect should apparently be taken into account (for example, periodically destroying the adsorbed layer by a pulsed increase in optical pumping \cite{bm-has}), but in a number of other applications this effect can be neglected.

In particular, it might be valid for solid-state resonance gravitational antenna of moderate sensitivity with an optical read out system, named as OGRAN detector \cite{bm-bag}. In this case, a partial improvement of the thermal contact between the detector and mirrors is possible \cite{bm-kvas}, but it cannot be complete in view of the necessity to preserve a high quality factor of the acoustic modes of mirrors \cite{bm-schw,bm-rana}. In addition the mechanical and optical stability of the mirror's multilayer reflecting coating under power laser pump becomes also the serious problem (in spite of a very low absorption coefficient) which has to be addressed experimentally. These objectives, under the general term of the ``interferometer with cryogenic mirrors'' is part of the researh program for a developing the third generation of gravitational wave antennas \cite{bm-punt,bm-ligo}.

In this article we report the particular result of our experimental study of behavior FP optical resonators with mirrors on substrate of different materials.

\section{Cryostat for cooling test bodies in optical experiments.}

To carry out the experiment, a pilot model of the OGRAN detector was manufactured \cite{bm-krys}. We used a solid state metallic (aluminum) bar with a tunnel along its central axis. Mirrors were attached at the ends of the bar, which composed the optical FP cavity. The model had the following parameters: length 200 mm, outer diameter 160~mm, tunnel diameter 20~mm, weight 8 kg. The detector was suspended inside the vacuum chamber of a special cryostat, having conductive thermal contact with its walls.

This cryostat consists of several cylindrical cavities, which are successively suspended inside each other on thin filaments of stainless steel. The external cavity is filled with liquid nitrogen, the next cavity is filled with liquid helium. The copper cooling chamber was suspended inside the helium cavity. Rapid pre-cooling of the chamber with the test object installed inside was carried out by passing liquid helium through a copper tube soldered along the entire length of the wall of the cooling chamber.

The vacuum system included a spiral oil-free pre-pump and a turbo molecular pump (Turbovac 350). At the degassing stage, the cryostat was evacuated up to a vacuum of at least $10^{-7}$~Torr.

This cryostat could operate in both flow and fill mode. In our experiments, we used a flow-through mode. Both chambers of the cryostat were filled with liquid nitrogen, while liquid helium was passed through a copper tube that enveloped the body of the cooling chamber The process of cooling the pilot model of mass 8 kg from the 77 to 5 K was carried out during about 3~hours and required about 30 litres of liquid helium.

A detailed description of the cryostat design and the suspension system of the test body inside the cooling chamber in the following paper \cite{bm-krys}.

The cryostat is equipped with one window for input and output of optical beam radiation. The temperature of the object was monitored by two commercial sensors from the Lakeshore company. One was mounted directly on the FP cavity mirror and the second was attached to the test body (i.e. bar of the pilot model).

\section{Substrate material for mirrors }

The choice of material for the mirror substrates for use gravitational-wave interferometers is primarily dictated by the requirement for small mechanical (acoustic) losses \cite{bm-rana}. This is due to the need to suppress the level of thermal (Brownian) noise, not only from the mirror's pendulum degree of freedom but also from their own acoustic modes. Experience has shown that some optical glasses, fused silica, sapphire, and a number of crystals, in particular silicon and fluorites (calcium fluoride CaF$_2$), satisfy this requirement \cite{bm-schw}. Optical glasses in astronomical instruments (telescopes) use materials with a low coefficient of thermal expansion (so called---zerodur , or Russian variant---sitall). In our test experiments, we used mirrors with substrates made of sitall, sapphire, and calcium fluoride (CaF$_2$). All of them have a wide transmission spectrum in the optical frequency range and they are able to withstand low temperatures without irreversible changes during warming. Experimental estimates of mechanical losses for materials are $\phi \sim 10^{-8}$ \cite{bm-schw} under cryogenic temperatures but the crystalline quartz has a jump to $10^{-5}$ in the temperature region $40 ^{o}K$. Although silicon is considered to be a promising material when operating at sub-nitrogen temperatures ($\sim 120$~K), it is opaque for the visible frequency range. Its use implies a transition to the wavelength $\lambda = 1.55\, \mu$m, which leads to a serious reconstruction of all elements of the optical path of the gravitational-wave interferometers. Our experiments were carried out in infrared light with $\lambda = 1.064 \mu$m. At this wavelength, all of the laser gravitational antennas work, including the OGRAN resonant optoacoustic detector \cite{bm-bag}.

Among the key parameters describing the physical properties of the substrate material and determining their behavior under low temperature conditions with high optical pump powers, a number of thermodynamic, mechanical, and optical characteristics are important \cite{bm-glen}. In particular,their heat capacity and thermal conductivity depend on the heating temperature and the uniformity of its distribution over the sample volume.

Young's modulus and thermo elasticity (coefficient of thermal expansion) affect the measurement of the mirror's deformation, t optical refractive index and its nonlinear component, and also the coefficients of optical reflection and absorption of incident radiation. The latter parameters are mainly determined by the quality and number of reflective layers of sprayed mirror coatings.

Our mirrors were manufactured (Polyus research institute of M.F.Stelmakh, Moscow, Russia $^3$) using multilayer dielectric interference coatings of alternating layers of oxides TiO$_2$/SiO$_2$ and Ta$_2$O$_5$/SiO$_2$, which were applied by ion beam spraying. The average parameters are: a scattering $K_s < 10$~ppm, absorption coefficient $K_a< 10$~ppm, a reflection coefficient $R \ge 99.998~$\%, laser strength $\sim (1-2)$~GW/cm$^2$.

It is useful here to provide a summary of parameters for the substrates in our study, see Table~\ref{tab:1}.

An important point in the experiment on cooling mirrors on substrates of these materials was to ensure sufficient thermal contact of the mirrors with the cryostat, so that the experiment could be carried out in a finite time.

Working through various mounting options led to the design shown in Fig.~\ref{fig:1}. The comments to the picture are as follows (the numbers and arrows mark particular details): the lower parts of the holder (2) are pressed (inserted on friction) into both bar ends of central tunnel of the PM (1), so that their planes are parallel to each other. The upper part of the holder (3) presses the inserted mirror (4) by means of screws (5) distributed along its perimeter. The clamping (press down) screws alternate with the spacer (wringing) screws, which allows the mutual optical tuning of the mirrors. In the upper part of holder of the input mirror, a groove is made to place the temperature sensor (6). The holder and screws are made of the same aluminum alloy (duralumin) as the body PM.

\section {Experimental setup and method of measurement}

The experimental setup consisted of the following main components, see Fig.~\ref{fig:2}: a) a cryostat containing a FP cavity with mirrors mounted on the ends of a cylindrical bar PM with a central tunnel; b) an optical node containing the laser of the EM-pump together with elements of a guide light and a photo detector plus the monitor, located on a separate desk; and c) control electronics unit. The configuration of the opto-electronic scheme and the principle of measurement are shown in Fig.~\ref{fig:2}.

The laser beam passes through a Faraday cell and an electro-optical modulator, and acquires a phase modulation with a frequency of 8 MHz. It is then directed to the studied FP-cavity. The beam reflected from the FP-cavity falls on the photo-detector, whose signal is fed to the monitor and to a spectrum analyzer. The photo-detector is formed from spherical and flat (input) mirrors.

In our experiment, we used a single mode tunable Nd:YAG laser with $\lambda = 1.064\, \mu$m with the maximum power up to P$\sim 1$W, which was developed in the Institute of Laser Physics SB RAS, Novosibirsk. Construction was based on the principle of a running wave ring laser \cite{bm-okhap}. Laser tuning in the low frequency region (0--5)~kHz was provided by piezo-ceramic stacks (slow and fast), which were attached to mirrors of the laser resonator.

In the experiment, two main integral characteristics of the FP cavity at the optical resonance were measured: $F$, finesse (or sharpness of the resonance); and $C$, contrast. By adjusting the laser frequency through the voltage applied to piezo-ceramics, one can tune the optical resonance of the FP cavity and measure its characteristics using the absorption curve in reflected light.

This curve appears on the oscilloscope when synchronizing the sweep speed with the rate of change of the laser frequency. However, the possibility of observing it depends on the laser's frequency stability and the width of the optical resonance. The laser device with pre-stabilization that we used \cite{bm-okhap} had a radiation line width of $\sim 10$kHz, which is less (but close) to the width of the FP resonance of $\sim 40$kHz and with finesse $F\sim 2000$. Therefore, we applied a calibration method to estimate the resonance width. The scale along the frequency axis (sweep axis) is set by the position of side modulation component in respect of the resonance frequency (their separation just equal the modulation frequency). The sweep was chosen so that the carrier and side-band were visible on the screen at the same time. The sharpness (finesse) was calculated as the ratio of the intermode interval (or free spectral range) $ f_{FSR} $ to the width of the resonance.

The contrast parameter $C$ gives the fraction of light involved in the interference (i.e. in optical resonance) and significantly affects the sensitivity of the FP standard in precision measurements.

According to the theory of a FP cavity, under absence of losses the intensity of reflected light equals zero at resonance frequency (i.e. the wave coming out from the cavity is compensated by the anti-phase wave reflected from the input mirror). This situation corresponds to the contrast parameter $C=1$ \cite{bm-black}.

In reality, the complete compensation does not exist due to losses (i.e. there is no zero intensity in reflected light on resonance). The measure of contrast is estimated by a relative depth of the dip in the absorption curve. Usually, an average value of the contrast under a satisfactory FP cavity tuning is on the order of $20-30$\%. The contrast essentially depends not only on the parameters of the mirrors themselves but also on the mode matching with the radiation structure (i.e. from the delicate accuracy of the resonance tuning). Temperature distortions of the mirrors obviously destroy the contrast.

\section{Measurements and results} 

Before placing the PM in the cryostat, the mirrors were mounted on it in a dust-proof box. The mirrors were fastened at the ends of the PM in special clips (holders) (Fig.~\ref{fig:1}). The compression of the mirror to the bottom of the holder was carried out by adjusting screws at four points along the periphery of the mirror. At the same time, the remaining gap between the clip and the lateral (cylindrical) surface of the mirror allowed it to react freely to radial thermal deformations. Only in one small area of the side surface did the mirror have a contact with the temperature sensor. When installing the mirrors, the initial tuning of the FP cavity using the auxiliary laser was carried out to find the FP resonances and the main TEM$_{00}$ mode. In this process, an optimal mutual axial orientation of the mirrors was selected by rotating them relative to the axis of the FP cavity to compensate for the initial deformation asymmetry (removing of the deformation of mode splitting) After this operation, the PM was moved to an open cryostat and hung on a special construction (the so-called $\Omega$-loop) with a loop coverage in the center of the PM. This design provided automatic maintenance of the thermal contact of the PM with the cryostat at all stages of cooling (for further details, see \cite{bm-krys}) A closed cryostat was placed on the atmosphere pumping out and it then underwent a cooling procedure. At the beginning of the optical measurements, fine tuning of the optical resonances of the FP-cavity was carried out by changing the angle of inclination of the incoming beam using an alignment platform in front of the cryostat. In the course of the experiment, the power of the radiation sent to the FP resonator could be changed stepwise from 450 mW to 5 mW using a set of absorbing filters. The measurements that we performed gave the following results.

\subsection*{a) Mirrors on sitall substrates.}

The material Sitall CO-115M or ``Astrositall'' refers to the category of ``zerodures'', which are materials with a low coefficient of thermal expansion ($1.5\cdot 10^{-7}$~1/$^o$C). 

The other solid-state characteristics are listed in Table~\ref{tab:1}. The optical properties are characterised by a small difference in the refractive indices of the extraordinary and ordinary rays $(n_{e} - n_{o}) \leq 0.001$ and, accordingly, the small birefringence parameter $d\leq 10$~NM/cm. Due to these properties, Astrositall is used to make the mirrors for large telescopes, including complex aperture synthesis mirrors.

The indicated characteristics of Sitall CO-115M are confirmed in the temperature range of -60$^o$C -- +60$^o$C. However, reliable data for cryogenic temperatures are unavailable. The results of our measurements of the evolution of the integral characteristics of a FP cavity with mirrors on substrates made of this sitall are shown in Fig.~\ref{fig:3}.

One can see a rather rapid drop in both parameters when the FP cavity is cooled. The finesse (red line) drops almost seven times as the temperature decreases by $100^o$~C. The contrast (blue line) weakens slightly slower, by three times when approaching the nitrogen temperature. Measurements at lower temperatures have already lost their meaning and have not been carried out. In the reverse course, during warming up, the parameters were restored (not shown), coming to the initial state with some hysteresis (a slight vertical shift for the same temperature). Thus, the observed changes were reversible.

\subsection*{b) Mirrors on sapphire substrates.}

According to its solid-state properties, sapphire is considered to be the most perfect material with a high Debye temperature and low internal losses that rapidly die out during cooling. Its modulus of elasticity is noticeably (3--4 times) larger than that of optical glasses and fluorites (i.e. less deformability). However, the tendency to double refraction is an order of magnitude higher than that of Sitall. The difference between the refractive indices of the extraordinary and ordinary rays $(n_e - n_o)\approx 0.01$. This creates certain problems when using sapphire mirrors in optical resonators at low temperatures. The results of measurements of the integral characteristics of a FP cavity with mirrors on sapphire substrates are shown in Fig.~\ref{fig:4}.

In practice, before the start of cooling, the resonance peak of the mode has a weak splitting, which can be roughly removed by mutual rotation of the mirrors.

However, upon cooling, the splitting arises again and grows approximately linearly in temperature. In the left-hand part of Fig.~\ref{fig:4}, the finesse and contrast values are substituted, which were estimated roughly by the same method, as long as the splitting value remained less than the peak width. Furthermore, these assessments are inadequate to the experiment task and were not made.

Fig.~\ref{fig:5} shows the splitting pattern at a temperature of $190^o$~K. The magnitude of the splitting reaches 120 kHz with a total coupled resonance width of $\sim 600$~kHz. Simulated (modeled) peaks are also represented here, which together give the observed picture (green curve - observed peak, blue curve - model peaks, red curve - synthesized curve). The points on the right-hand side of the figure give a rough estimate of the contrast, if one ignores the presence of splitting.

\subsection*{c) Mirrors on calcium fluoride substrates.}

Calcium fluoride is a widespread material in IR spectroscopy in the wavelength range $(0.15-9) \mu$m. The CaF$_2$ crystal has high mechanical strength (withstands high pressure) and is non-hygroscopic. It is believed that it is sensitive to thermal effects (although the melting point is rather high 1~418~$^o$C). At room temperature it is optically isotropic. The difference between the refractive indices of the extraordinary and ordinary rays is very small $(n_e - n_o) \lesssim 0.001$.

Our measurements of the FP cavity with mirrors on calcium fluoride substrates showed that its integral characteristics (finesse and contrast) are practically kept in a wide range of temperatures, from room to helium. A slight drop in the contrast 20\% was observed during the passage beyond the mark of $60^o$~K. (It should be noted that the thermal conductivity of calcium fluoride in the region from 60K to 40K increases from $\sim$1~W/cm$\cdot$K up to $\sim$10~W/cm$\cdot$K \cite{bm-glen}. Perhaps this could change the deformation balance in the mirror substrate and destroy the mode matching.)

This is reflected in Fig.~\ref{fig:6}, where the cooling curves for the two optical pump powers 10~mW and 350~mW are presented. When warming up, the contrast value was restored with a small hysteresis, which is the same as for the mirrors with other substrates studied above.

\section{Discussion and conclusions}

As mentioned in the introduction, optical resonators FP are used for pre-stabilization of lasers in schemes of optical frequency standards and for registration of small opto-acoustic perturbations in gravitational antennas. The known PDH technique \cite{bm-drever,bm-black} is applied, where the frequency stabilization of the laser is due to the connection with the reference optical resonator. The effectiveness of this scheme is determined by the amount of sharpness (finesse) and contrast of this resonator, which should be large. The desire to suppress electromagnetic and mechanical thermal noise forces us to consider FP resonators that are cooled to cryo-temperatures. However, the possibility of preserving their optical properties requires an experimental investigation.

This paper presents a study of the temperature evolution of the FP cavity for three types of mirrors. The main result is a demonstration of the suitability of mirrors on calcium fluoride substrates. The flaws of sitall and sapphire based mirrors were also discovered.

The reasons for the fall in the quality of the FP of resonators with sitall mirrors are not completely clear. Given that their optical nonlinearity is small, we can assume a deterioration in the optical qualities of such mirrors is due to thermo-mechanical effects. However, this hypothesis requires additional research.

Optical nonlinearity (birefringence) affects sapphire mirrors. It is obvious that with a large frequency splitting of resonances that preserves the high finesse, the use of the PDH technique may also be applied to a separate isolated peak. However, this does not prevent the loss of contrast because the optical power is spread over all of the peaks.

The obtained results are directly important for the implementation of the cryogenic version of the OGRAN \cite{bm-kvas} gravitational detector. This setup uses LMA mirrors \cite{bm-lma} on quartz substrates with a very high finesse value of the reference resonator $\sim$~30~000. Quartz birefringence is close to that of sapphire. With this increased finesse, the peak splitting will be large enough to use the PDH technique for the separate one.

\section*{Acknowledgement}

The authors would like to express their gratitude to the members of OGRAN team from LPI SB RAS (Novosibirsk), particularly: Bagaev S.N., Skvortsov M.N. and Kvashnin N.N. for their help in instrumentation and many fruitful discussions. This work was supported by national grant RSCF~17-12-01488 and partly by grant RFBR~17-02-00492.


\newpage
\begin{table}[!h]
\caption { Substrate constants.} \label{tab:1} 
\begin{tabular}{|c|c|c|c|}\hline
Material & CaF$_2$ & Sapphire & Sitall CO-115M \\ \hline
Refractive index $n_e$ & 1.429 & 1.746 & 1.538 \\ \hline
Refractive index $n_o$ & 1.429 & 1.754 & 1.536 \\ \hline
Young modulus, GPa & 75.8 & 335 & 92 \\ \hline
Thermal expansion coefficient, $10^{-6}$(1/K) & 5.6 & 18,5 & 0.15 \\ \hline
Thermal conductivity, (W/M$\cdot$K) & 27.21 & 9.71 & 1.99 \\ \hline
Specific heat, (J/(Kg$\cdot$K) & 419 & 854 & 920 \\ \hline
density, (g/cm$^3$) & 3.97 & 3.18 & 2.46 \\ \hline
\end{tabular}
\end{table}

\newpage
\begin{figure}[h]
\caption{Mounting of mirror holder to the body of pilot model: (1) pilot model body, (2) holder lower part, (3) holder upper part, (4) mirror, (5) screws, and (6) temperature sensor.}
\label{fig:1}
\end{figure}

\newpage
\begin{figure}[h]
\caption {Experimental setup configuration.}
\label{fig:2}
\end{figure}

\newpage
\begin{figure}[h]
\caption{Evolution of the finesse and contrast for FP-cavity with mirrors on sitall during the cooling process.}
\label{fig:3}
\end{figure}

\newpage
\begin{figure}[h]
\caption{Evolution of the finesse and contrast for FP-cavity with mirrors on sapphire during the cooling process}
\label{fig:4}
\end{figure}

\newpage
\begin{figure}[h]
\caption{Splitting of optical resonance with cooled sapphire mirrors: T=190$^o$K, splitting of peaks $\sim$120kHz, width of coupled resonance $\sim$600Hz (green - observed peak, blue - model peaks, and red - synthesized curve).}
\label{fig:5}
\end{figure}

\newpage
\begin{figure}[h]
\caption{Evolution of the finesse and contrast for FP-cavity with mirrors on calcium fluoride during the cooling process.}
\label{fig:6}
\end{figure}

\newpage
\listoftables
\listoffigures

\end{document}